\documentclass{appolb}
\usepackage{epsfig}
\usepackage{axodraw-1}
\def\lsim{\mathrel{\raise.3ex\hbox{$<$\kern-0.8em\lower1ex\hbox{$\sim$}}}}
\def\gsim{\mathrel{\raise.3ex\hbox{$>$\kern-0.7em\lower1ex\hbox{$\sim$}}}}

\def\eg{{\it e.g.}}
\def\etc{{\it etc.}}

\begin{document}
\title{Color-octet scalars at the LHC\thanks{Presented by J. Kalinowski at the Epiphany Conference
on Hadron Interactions at the Dawn of the LHC, 5-7 January 2009, Cracow, Poland.}%
}
\author{S.~Y.~Choi$^1$, M.~Drees$^{2}$, J.~Kalinowski$^3$,
        J.~M.~Kim$^2$, E.~Popenda$^4$ and P.~M.~Zerwas$^{4,5}$
\address{$^1$ Dept. Physics and RIPC, Chonbuk National University,
               Jeonju 561-756, Korea \\
          $^2$ Phys. Inst. and Bethe CPT, U.~Bonn, D-53115 Bonn, Germany \\
          $^3$ Inst.\ Theor.\ Physics, U.~Warsaw, PL-00681 Warsaw, Poland \\
          $^4$ Inst.~Theor.~Physik E, RWTH Aachen, D-52074 Aachen, Germany \\
          $^5$ Deutsches Elektronen-Synchrotron DESY, D-22603 Hamburg, Germany} }
               \maketitle
\begin{abstract}
Elements of the phenomenology of color-octet scalars (sgluons), as predicted in the hybrid N=1/N=2 supersymmetric model, are discussed in the light of forthcoming experiments at the CERN Large Hadron Collider.
\end{abstract}
\PACS{12.60.Jv, 14.80.Ly}

\section{Motivation}
After many years of preparations and constructions, the Large Hadron Collider at CERN will start its operation soon.
One of the most important questions the LHC should provide an answer to is related to the nature
of the electroweak gauge symmetry breaking. In the Standard Model (SM) it is achieved by adding a doublet of Higgs fields and arranging the parameters
such that the symmetry is spontaneously broken. Although the SM is extremely successful in describing the experimental data,
there are many arguments suggesting that it cannot be the ultimate theory,  \eg\ the SM Higgs sector is unnatural --
cannot explain why the electroweak scale is so small with respect to the Planck scale; the SM cannot account
for the matter-antimatter asymmetry of the Universe, nor for the dark matter \etc\
In fact, the above shortcomings point towards a new physics at the TeV scale.

Among many propositions for solving part of these problems in the physics area beyond the Standard Model,
supersymmetry (SUSY) is generally considered most elegant and respected.  Not invented or
designed to,  it can accommodate or explain some of the outstanding problems of the SM:
stabilizing the gap between electroweak and Planck scale, unifying the gauge couplings, inducing radiative
electroweak symmetry breaking. It
provides a candidate for dark matter (DM), offers new ideas on the
matter-antimatter asymmetry, while its unique mathematical structure can provide a link to physics
at the GUT/Planck scale and, in local form, paving the path to gravity.

In the simplest N=1 supersymmetric extension of the SM, each SM particle is paired with  a sparticle that differs in
spin by half a unit. Since none of the sparticles has been seen so far, the  supersymmetric Lagrangian must be supplemented by SUSY breaking terms that keep unseen superpartners out of experimental reach while retaining renormalizability and maintaining perturbatively stable hierarchy of scales.
Experimental constraints, mainly from flavor and Higgs physics, limit the allowed parameter space and play an increasingly restrictive role in building models of SUSY breaking.

However, the successes of supersymmetry do not rely on its simplest realization. In fact, non-minimal realizations may ameliorate the flavor problem. For example, Dirac gauginos (in contrast to Majorana in the MSSM) forbid some of the couplings and often lead to additional suppression of contributions from loops with gauginos
in flavor-changing processes. Such scenarios can be based on D-term supersymmetry breaking models \cite{fnw,npt} or continuous R-symmetries \cite{rsymm}.

A Dirac gaugino requires additional fermionic degrees of freedom. They can be provided by adding a chiral super-multiplet in the adjoint representation of the gauge group \cite{Fayet}.
Here we will present elements of the phenomenology of the scalar partner of the Dirac gluino (sgluon), as worked out in a recent paper~\cite{our}
[and confronted with Ref.~\cite{TT}].

\section{Short introduction to the N=1/N=2 hybrid model}

In the MSSM gluinos are Majorana fields with two degrees of freedom to match gluons in the color-octet vector super-multiplet.
To provide the two additional degrees of freedom for Dirac fields, the usual N=1 gluon/gluino
vector super-multiplet $\hat{g}=\{g_\mu, \tilde{g} \}$ may be paired with an additional N=1 color-octet chiral super-multiplet $\hat{g}'=\{\tilde{g}',\sigma\}$ of extra gluinos and scalar
$\sigma$ fields to a vector hyper-multiplet of N=2 supersymmetry. [Similarly, the electroweak sector, not to be discussed here,  is supplemented by additional SU(2)$_L$ and U(1)$_Y$ super-multiplets.] The N=2 mirror
(s)fermions are assumed to be very heavy in order to avoid chirality problems. This hybrid
model expands to N=2 only in the gaugino sector~\cite{CDFZ}.

\subsection{The gluino sector}
Standard MSSM $\tilde{g}$ and new gluinos $\tilde{g}'$ couple minimally to the gluon field
\begin{equation}
    {\mathcal{L}}_{\rm SQCD} \ni g_s {\rm Tr}\,
    ( \overline{\tilde{g}} \gamma^{\mu} [{g}_{\mu}, {\tilde{g}}] +
          \overline{\tilde{g}'} \gamma^{\mu} [{g}_{\mu}, {\tilde{g}'}] ) \,,
  \label{eq:gluino}
\end{equation}
where $g_s$ denotes the QCD coupling, the fields being color-octet matrices (\eg\ ${g}_\mu =
\frac{1}{\sqrt{2}} \lambda^a g^a_\mu$ with the Gell-Mann matrices $\lambda^a$);
$\tilde{g}$  and $\tilde{g}'$  are two 4-component Majorana spinor fields.  Quark and squark fields interact only with the standard gluino,
\begin{equation} \label{eq:qcd-yuk}
   \mathcal{L}_{\rm SQCD} \ni
   -  g_s (\,\overline{q_L}  \tilde{g} \, \tilde{q}_L
                 - \overline{q_R}  \tilde{g} \, \tilde{q}_R
                 + {\rm h.c.}) \,,
\end{equation}
since only their hyper-multiplet partners (assumed to be heavy) couple to $\tilde{g}'$, as required by N=2 supersymmetry.

Soft supersymmetry breaking generates masses for the gluino fields $\tilde{g}$
and $\tilde{g}'$,
\begin{equation}
   {\mathcal{M}}_g =\left(\begin{array}{cc}
   M'_3 & M^D_3 \\
   M^D_3 & M_3
\end{array}\right).
\label{eq:gluinomass}
\end{equation}
Diagonal terms are induced by the individual Majorana mass parameters $M_3$ and $M_3'$ while an
off-diagonal term corresponds to the  Dirac mass. Diagonalization gives rise to two Majorana mass eigenstates, $\tilde{g}_1$ and $\tilde{g}_2$,
with masses $m_1$ and $m_2$. There are two limiting cases of interest: in the limit $M_3' \to \pm
\infty$ the standard MSSM gluino is recovered;  in
the limit of vanishing Majorana mass parameters $M_3$ and $M_3'$ with
 $M^D_3\neq 0$, the mixing is
maximal and the two Majorana gluino states are paired
to a Dirac state,
\begin{equation}
\tilde{g}_D=\tilde{g}_R+\tilde{g}'_L.
\end{equation}

\begin{figure}\begin{center}
\epsfig{figure=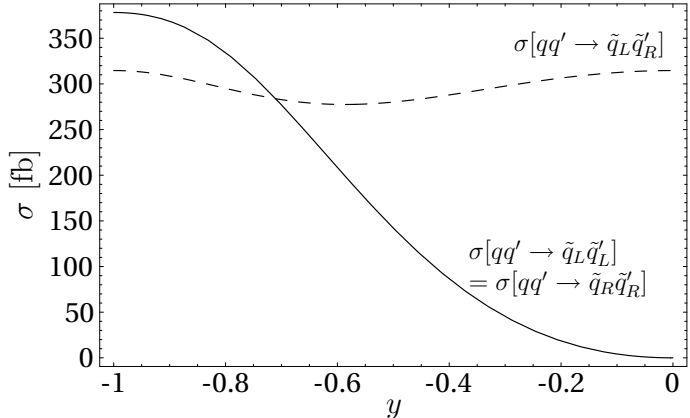, width=9cm}
\vspace{-2ex}
\caption{\it Partonic cross sections for different-flavor squark production as a
function of the Dirac/Majorana control parameter~$y$.}
\label{fig:D/M}\end{center}
\end{figure}

Dirac gluinos are characteristically different from Majorana gluinos; for detailed discussion we refer to \cite{CDFZ}. Here we present one example from~\cite{CDFZ} in Fig.~\ref{fig:D/M}, where the partonic cross sections for different-flavor squark production $qq'\to\tilde{q}\tilde{q}'$ mediated by the gluino t-channel exchange are plotted as a
function of a Dirac/Majorana control parameter~$y$, assuming partonic center-of-mass energy $\sqrt{s}=2000$ GeV,
$m_{\tilde q} = 500$ GeV and $m_{\tilde{g}_1} = 600$ GeV. Here $M'_3=y\, M^D_3/(1+y)$, $M_3=-y\, M^D_3$ with $M^D_3=m_{\tilde{g}_1}$ kept fixed along $y\in [-1,0]$. The parameter $y$ allows for a continuous transition from $y=-1$, where the MSSM limit is reached with one Majorana gluino (the second being infinitely heavy), to $y=0$ that corresponds
to two degenerate Majorana fields combined to a Dirac gluino.
For equal $\tilde q_L$ and $\tilde q_R$ masses, $\sigma(qq' \to \tilde q_L \tilde q_L')= \sigma(qq' \to \tilde q_R \tilde q_R') $ are non-zero in the Majorana limit
but vanish in the Dirac, while
$ \sigma(qq' \to \tilde q_L \tilde q_R')$ reaches the same value in both limits.

\subsection{The sgluon sector}

The new gluinos are accompanied by a color-octet complex scalar field $\sigma$.
In the simplest realization, in parallel to the degenerate Majorana gluinos combined
to the Dirac gluino, we assume that the real and imaginary components of the scalar field $\sigma$ are degenerate with the (sgluon) mass denoted by $M_\sigma$~\cite{our}. At tree level, apart from the $\sigma\sigma^*g$ and $\sigma\sigma^*gg$ couplings as required by the gauge symmetry,
the sgluons couple to the gluino pair via the Yukawa-type interaction
\begin{eqnarray}
{\cal L}_{\tilde{g}_D\tilde{g}_D\sigma}
   &=& -\sqrt{2} i\, g_s\, f^{abc}\, \overline{\tilde{g}^a_{DL}}\,
        \tilde{g}^b_{DR}\, \sigma^c
       +{\rm h.c.},\label{eq:sglgl}
\end{eqnarray}
where $f^{abc}$ are the SU(3)$_C$ structure constants.
When supersymmetry is broken spontaneously, the Dirac gluino mass generates a scalar coupling between $\sigma$ and squark pair~\cite{fnw}
\begin{equation} \label{eq:sqq}
{\cal L}_{\sigma\tilde{q}\tilde{q}} =  -
 g_s M_3^D\,  \sigma^a \frac{\lambda^a_{ij}} {\sqrt{2}}
  \sum_q \left( \tilde q_{Li}^* \tilde q_{Lj} - \tilde q_{Ri}^* \tilde q_{Rj}
  \right) + {\rm h.c.}\,,
\end{equation}
where $L$ and $R$ squarks contribute with
opposite signs, as demanded by the general form of the super-QCD $D$-terms.

Although sgluons are $R$-parity even,  the above couplings imply that tree-level single production
of $\sigma$'s in gluon-gluon or quark-antiquark collisions is not possible. Only $\sigma$ pairs can be produced in gluon collisions as well as in $q \bar q$ annihilation.

Since sgluons couple to squarks and gluinos, loop-induced couplings to the SM fields are of interest. However,
even at the one-loop level gluino loops do not contribute to the $\sigma gg $
coupling as a consequence of Bose symmetry since the coupling is even in momentum space but odd,  $\sim f^{abc}$,
in color space.\footnote{Actually, the coupling of the octet sgluon to any number
of gluons via the gluino loop is forbidden  since the sgluon couples only to two different Majorana gluinos, eq.~(\ref{eq:sglgl}), while gluons always couple to the same pair, eq.~(\ref{eq:gluino}).}
\begin{figure}[t]
\begin{center}
\begin{picture}(300,60)(0,70)
\Text(0,95)[r]{$(a)$}
\Text(15,95)[c] {$\sigma$}
\Text(100,95)[c] {$\tilde{q}$}
\DashLine(25,95)(50,95){3}
\DashLine(50,95)(90,115){3}
\DashLine(90,75)(50,95){3}
\DashLine(90,115)(90,75){3}
\Text(125,115)[l]{ $g$}
\Text(125,75)[l]{ $g$}
\Gluon(90,115)(120,115){2}{5}
\Gluon(90,75)(120,75){2}{5}
\Text(150,95)[r]{ $\sigma$}
\Text(200,100)[c]{ $\tilde{q}$}
\DashLine(155,95)(180,95){3}
\DashArrowArcn(200,95)(20,180,0){3}
\DashArrowArcn(200,95)(20,0,180){3}
\Gluon(220,95)(250,115){2}{4}
\Gluon(220,95)(250,75){2}{4}
\Text(255,115)[l]{ $g$}
\Text(255,75)[l]{ $g$}
\end{picture}
\begin{picture}(300,60)(0,70)
\Text(0,95)[r]{$(b)$}
\Text(15,95)[r]{ $\sigma$}
\Text(70,115)[c]{ $\tilde{g}_D$}
\Text(95,95)[l]{ $\tilde{q}$}
\DashLine(25,95)(50,95){3}
\Line(50,95)(90,115)
\Line(90,75)(50,95)
\DashLine(90,75)(90,115){3}
\Text(125,115)[l]{ $q$}
\Text(125,75)[l]{ $\bar{q}$}
\ArrowLine(90,115)(120,115)
\ArrowLine(120,75)(90,75)
\Text(150,95)[r]{ $\sigma$}
\Text(200,115)[c]{ $\tilde{q}$}
\Text(225,98)[l]{ $\tilde{g}_D$}
\DashLine(155,95)(180,95){3}
\DashLine(180,95)(220,115){3}
\DashLine(220,75)(180,95){3}
\Line(220,75)(220,115)
\ArrowLine(220,115)(250,115)
\ArrowLine(250,75)(220,75)
\Text(255,115)[l]{ $q$}
\Text(255,75)[l]{ $\bar{q}$}
\end{picture}
\end{center}
\caption{\it Generic diagrams for  the effective $\sigma gg$ (a) and $\sigma q\bar{q}$ (b) vertices  with
             squark/gluino loops.}
\label{fig:loops}
\end{figure}
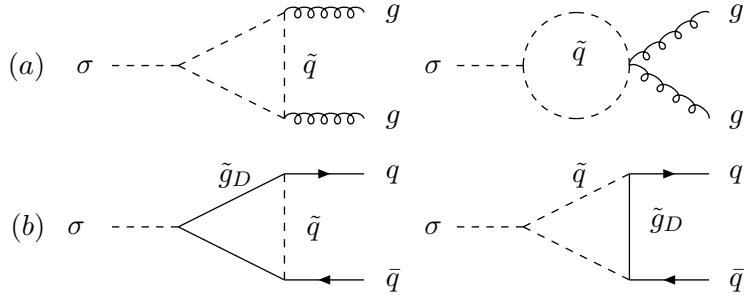
On the other hand,  triangle diagrams involving squark/gluino lines (Fig.~\ref{fig:loops}) generate $\sigma gg$ and
$\sigma q \bar q$ couplings.  The
interaction Lagrangian however, eq.~(\ref{eq:sqq}), implies that all $L$- and $R$-squark contributions
to the couplings come with opposite signs so that they cancel each other for
mass degenerate squarks. In addition, the quark-antiquark coupling is
suppressed by the quark mass as evident from general chirality rules.

\section{The phenomenology of sgluons at the LHC}
Sgluons have a large color charge and thus might be copiously produced at the LHC.
First we discuss their decay modes, and then production processes at the LHC.

\subsection{Sgluon decays}
Sgluons  will decay into different channels that include gluinos, squarks, gluons and quarks.
At tree level the $\sigma$ particles can decay to a pair of Dirac gluinos
${\tilde{g}}_D$ or into a pair of squarks,
\begin{eqnarray} \label{eq:tree1}
\Gamma [\sigma \to \tilde{g}_D {\bar{\tilde{g}}}_D ]
   &=& \frac{3 \alpha_s M_{\sigma}}{4}
        \beta_{\tilde{g}}\, (1+\beta^2_{\tilde{g}})\,, \\
\Gamma[ \sigma \to \tilde{q}_a \tilde{q}_a^*] &=& \frac {\alpha_s
   |M_3^D|^2} {4 M_\sigma} \beta_{\tilde{q}_a}\,, \label{eq:tree2}
\end{eqnarray}
where $\beta_{\tilde{g},\tilde{q}_a}$ are the velocities of $\tilde{g},\tilde{q}_a$  ($a=L,R$). If $M_{\sigma} < 2 M_{{\tilde{g}}_D}, 2 m_{\tilde q}$, one or both of these sparticles can be virtual.  Squarks and gluinos  will further cascade to SM particles and LSP.

Loop-induced couplings will mediate decays into gluon or quark-antiquark pairs
\begin{eqnarray}
\label{loop-gg}
\Gamma(\sigma \rightarrow g g)
  &=& \frac {5 \alpha_s^3} {384 \pi^2} \frac{|M_3^D|^2} {M_\sigma}
      | \sum_q [ \tau_{\tilde{q}_L} f(\tau_{\tilde{q}_L})
                         - \tau_{\tilde{q}_R} f(\tau_{\tilde{q}_R}) ]|^2\,,\\
\Gamma(\sigma \rightarrow q \bar q)
   &=& \frac{9 \alpha_s^3} {128 \pi^2} \frac{|M_3^D|^2 m_q^2}{M_\sigma}\,
     \beta_q
     [ (M^2_\sigma-4 m_q^2) |{\cal I}_S|^2
           +M^2_\sigma \, |{\cal I}_P|^2 ]\,. \label{loop-qq}
\end{eqnarray}
In eq.(\ref{loop-gg}), $\tau_{\tilde{q}_{L,R}} = 4 m^2_{\tilde q_{L,R}} / M^2_\sigma$ and
$f(\tau)$ is the standard  function from a squark circulating in the  loop~\cite{MS}. In eq.(\ref{loop-qq}),
 the effective scalar ($S$) and pseudoscalar ($P$)
couplings  take the form ($\alpha=S,P$)
\begin{eqnarray} \label{I_loop}
{\cal I}_\alpha = \int_0^1 dx \int_0^{1-x} dy [w_\alpha (  C^{-1}_L - C^{-1}_R )
      +z_\alpha (D^{-1}_L - D^{-1}_R) ]
       \,,
\end{eqnarray}
where $w_S=1-x-y$, $w_P=1$, $z_S=(x+y)/9$, $z_P=0$,  and the squark/gluino denominators
are ($a= L,R$) \\[3mm]
$C_a = (x+y) |M_3^D|^2 + (1-x-y) m^2_{\tilde q_a} - x y M^2_\sigma
 - (x+y) (1-x-y) m_q^2$,\\[1mm] $
D_a = (1-x-y) |M_3^D|^2 + (x+y) m^2_{\tilde q_a}
- x y M^2_\sigma - (x+y) (1-x-y) m_q^2$.\\[-1mm]

If $\tilde
q_L$ and $\tilde q_R$ of a given flavor mix, the subscripts $L,R$ in equations above have to
be replaced by $1,2$ labeling the squark mass eigenstates, and the
contribution from this flavor is suppressed by the
factor $\cos ( 2 \theta_q)$. Note that, because of the chirality structure, decays to light quarks are suppressed by the quark mass and that  both loop-induced decays are absent if $L$ and $R$ squarks are degenerate

The ordering between the tree-level and loop-induced
decay modes depends on the values of various soft
breaking parameters.
So long as gluino-pair decay channels are shut kinematically, even for small $L$-$R$ squark mass splitting the sgluon decays into two gluons, and to a $t \bar t$ pair if
kinematically allowed, always dominate over tree-level off-shell
four--body decays $\sigma \rightarrow \tilde g q \bar q \tilde \chi$  and  $\sigma \rightarrow q \bar q \tilde \chi \tilde \chi$. Increasing the gluino mass increases the $\sigma \tilde q
\tilde q^*$ coupling.  As a result, the partial width into two gluons (due
to pure squark loops) increases, while the $t \bar t$ partial width
(due to mixed squark--gluino loops)  decreases since the increase of the $\sigma \tilde q \tilde q$ couplings is
over--compensated by the gluino mass dependence of the propagators.
Of course,  the tree--level, two--body decays of eqs.~(\ref{eq:tree1},\ref{eq:tree2}) will dominate if they are kinematically allowed. Well above all thresholds the partial width into gluinos always
dominates: it grows $\propto M_\sigma$ while the partial width into
squarks asymptotically scales like $1/M_\sigma$ since the supersymmetry breaking $\sigma \tilde q \tilde q^*$ coupling has mass
dimension 1, while the supersymmetric $\sigma \tilde g \bar{\tilde g}$
coupling is dimensionless.

\begin{figure}[t]
\vskip 1mm
\begin{center}
\rotatebox{270}{\epsfig{figure=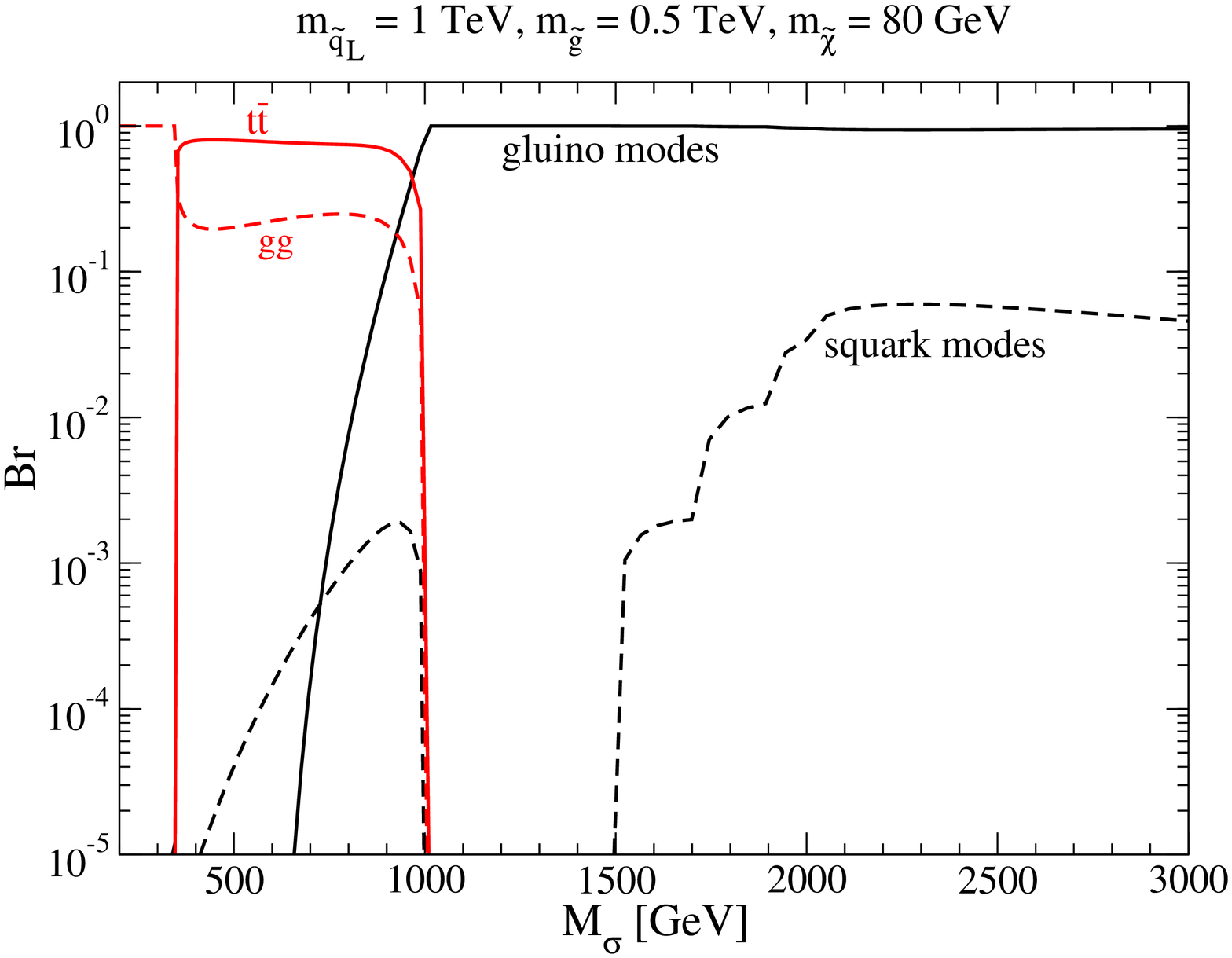, width=5.4cm,height=6cm}}
\rotatebox{270}{\epsfig{figure=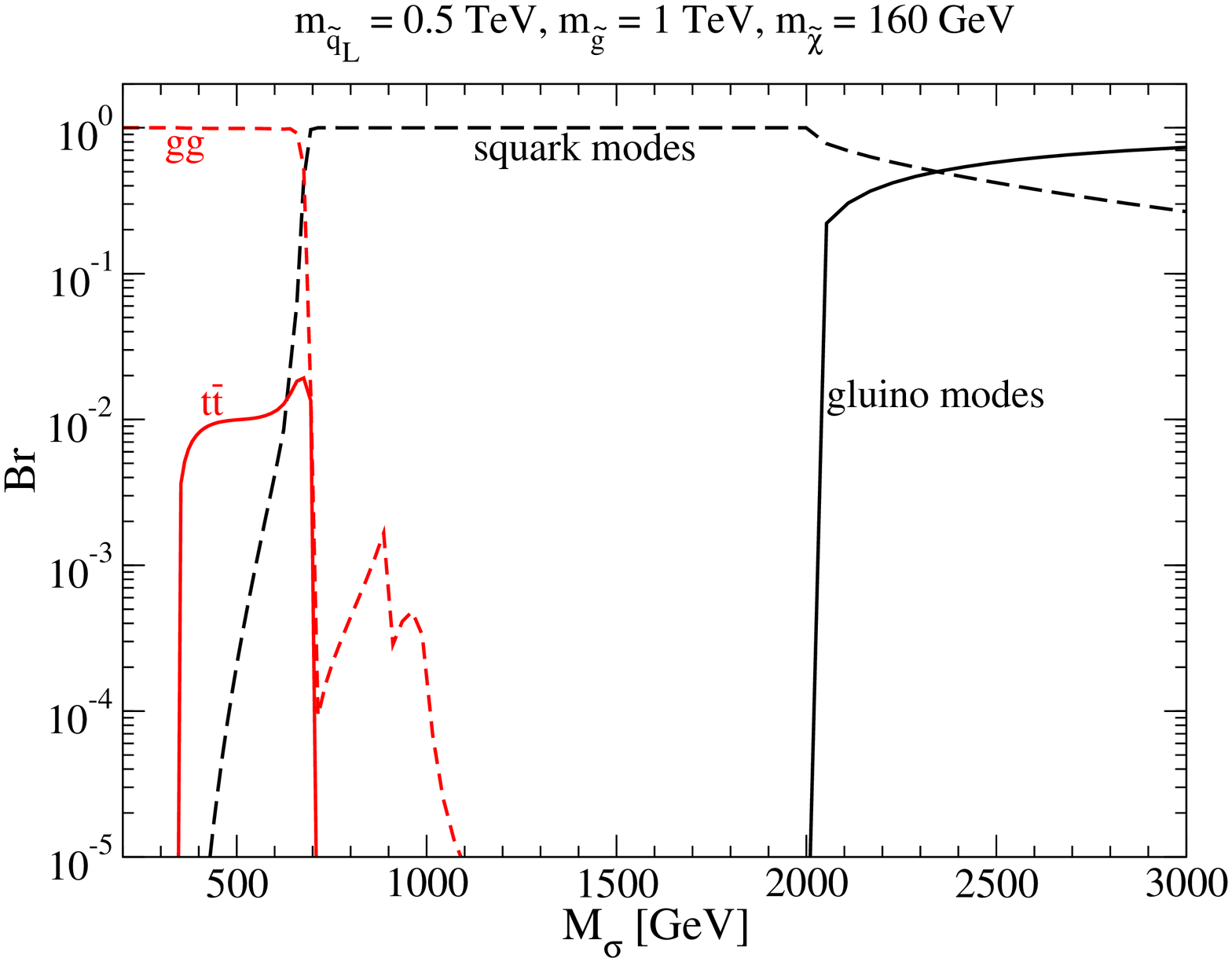, width=5.4cm,height=6cm}}
\caption{\it Branching ratios for $\sigma$ decays, for $m_{\tilde q_L}$=2$
  m_{\tilde g}$=1~TeV (left) and $m_{\tilde g}$ = 2$m_{\tilde q_L}$=1~TeV
  (right). }
\label{fig:br}\end{center}
\end{figure}

The above qualitative features can be  seen in Fig.~\ref{fig:br}, where the branching
ratios for $\sigma$ decays are plotted for two different squark/gluino mass hierarchies. Moderate mass
splitting between the $L$ and $R$ squarks of the five light flavors, and
somewhat greater for soft breaking $\tilde t$ masses have been assumed: $m_{\tilde q_R} = 0.95
m_{\tilde q_L}, \, m_{\tilde t_L} = 0.9 m_{\tilde q_L},\, m_{\tilde t_R} = 0.8
m_{\tilde q_L}$, the off--diagonal element of the squared
$\tilde t$ mass matrix $X_t = m_{\tilde q_L}$; and the gluino is a pure Dirac state, i.e. $m_{\tilde g} = |M_3^D|$.

\subsection{Sgluon production and signatures at the LHC}
\begin{figure}[t]
\begin{center}
\epsfig{figure=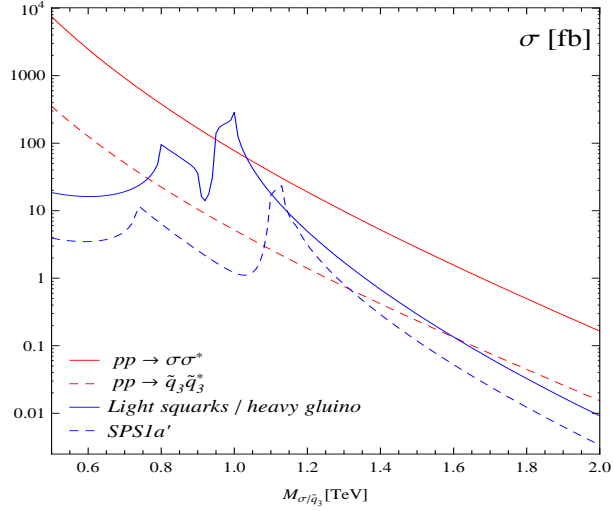, height=6.8cm, width=8cm}
\caption{\it Cross sections for $\sigma$-pair [and $\tilde{q}_3$-pair]
         production (red lines), as well as for single $\sigma$ production (blue lines),
         at the LHC. The solid blue curve is for the mass parameters as in Fig.3 (right), while the dashed blue for the mSUGRA benchmark point SPS1a$'$.}
\label{fig:lhcsigma}
\end{center}
\end{figure}

Sgluons can be produced singly in gluon-gluon collisions
via squark loops; the production via $q\bar q$ annihilation is negligible for light incoming quarks. The partonic cross section is given by
\begin{equation} \label{sig_gg}
{\hat{\sigma}} [ gg \to \sigma ] = \frac{\pi^2} {M^3_\sigma} \Gamma(\sigma \to
gg)\,,
\end{equation}
where the partial width for $\sigma \to gg$ decays is given in
eq.$\,$(\ref{loop-gg}).
The resulting cross section for single $\sigma$ production at the LHC is shown
by the blue curves in Fig.$\,$\ref{fig:lhcsigma} (with the LO CTEQ6L parton
densities \cite{CTEQ}). The solid curve has been calculated for the parameter
set of the right frame of Fig.$\,$\ref{fig:br}, while the dashed one is for  the benchmark point SPS1a$'$ \cite{sps}.  Since SPS1a$'$ has a
somewhat smaller gluino mass (interpreted here  as a Dirac mass)
it generally leads to smaller cross sections for single $\sigma$ production.

The signatures for single $\sigma$ production are potentially very exciting.
However, the 2-gluon decay channel must be discriminated from the large SM background. On the other hand, large production cross section in the gluon fusion would imply diminished decay rates  to other channels, which  in addition do not allow   a  direct reconstruction of $M_\sigma$. Detailed experimental simulations are needed to see if the single $\sigma$ production can be
detected as a resonance above the SM and MSSM backgrounds.

Sgluons can also be pair-produced in $q\bar q$ and $gg$ processes,
\begin{eqnarray}
\sigma [q\bar{q} \to \sigma\sigma^{\ast} ] &=& \frac{4 \pi \alpha_s^2}{9s}
     \,\beta^3_\sigma \,, \\
\sigma [gg \to \sigma\sigma^{\ast} ]
     &=& \frac{15 \pi \alpha_s^2\beta_\sigma}{8s}
     [ 1 + \frac{34}{5}\, \frac{M_\sigma^2}{s}-\frac{24}{5}
     (1-\frac{M_\sigma^2}{s})\frac{M_\sigma^2}{s}\,
     L_\sigma ]\,,
\end{eqnarray}
where $\sqrt{s}$ is the
invariant parton-parton energy, $M_{\sigma}$ and $\beta_\sigma$ are the mass and center-of-mass velocity of the
$\sigma$ particle, and $L_\sigma =\beta^{-1}_\sigma
     \log(1+\beta_\sigma)/(1-\beta_\sigma)$.
The cross section for $\sigma$-pair production at LHC, $pp \to \sigma
\sigma^\ast$, is shown by the solid red curve in Fig.$\,$\ref{fig:lhcsigma}
for the $\sigma$-mass range between  500 GeV and 2 TeV. With
values from several picobarn downwards, a sizable $\sigma\sigma^\ast$ event
rate can be generated. As expected, due to large color charge of the sgluon, the $\sigma\sigma^\ast$ cross section exceeds stop or sbottom-pair
production (red dashed line), mediated by a set of topologically equivalent Feynman diagrams,
by more than an order of magnitude. In the scenario of the right frame of  Fig.~\ref{fig:br}, the single $\sigma$ cross section can exceed the $\sigma$-pair
production cross section for $M_\sigma \sim 1$ TeV; taking $m_{\tilde q} \simeq 2 |M_3^D|$, as in the left frame of Fig.~\ref{fig:br},
lead to a very small single $\sigma$ production cross section. In general one cannot simultaneously have a large $\sigma(pp \to \sigma)$ and a large ${\rm Br}(\sigma \to t \bar t)$.

With the exception of $\sigma \rightarrow g g$ decays, all the $\sigma$ decay modes  give rise to signatures that should be easily detectable. Most spectacular signature results   from $\sigma
\rightarrow \tilde g \tilde g$ decay, each $\sigma$ decaying into at least
four hard jets and two invisible neutralinos as LSP's. $\sigma$-pair
production then generates final states with a minimum of eight jets with high sphericity and four LSP's.
In such events the transverse momenta of the hard jets produced and
the vector sum of the
momenta of the four ${\chi}^0_1$ in the final state, which determines
the measured missing transverse momentum $p_T$,  are
markedly different from the corresponding MSSM gluino or squark production with the same mass
configurations~\cite{our}.

Other interesting final states  are
four-top states $\tilde t_1 \tilde t_1 \tilde t_1^* \tilde t_1^*$,
which can dominate if $m_{\tilde q} \lsim m_{\tilde g}$ and $L$-$R$ mixing is significant
in the stop sector, and $\tilde q \tilde q^* \tilde g \tilde g$ if $M_{\sigma} > 2 m_{\tilde g} \gsim 2 m_{\tilde q}$.
These channels also lead to four LSPs in the final state, plus a large number
of hard jets. On the other hand, the $t t \bar t \bar t$ final state, which
can be the dominant mode if the two--body decays into squarks and gluinos are
kinematically excluded, might allow the direct kinematic reconstruction of
$M_\sigma$. Observation in addition $t c \bar t \bar c$ final state might indicate a substantial mixing in the up-type squark sector.

\section{Summary}
Models with Dirac gauginos offer an interesting alternative to the MSSM scenario.
Embedded into theories of extended supersymmetries, they
predict the presence of scalar particles in the adjoint representation of the gauge groups.
The color-octet scalars, sgluons, (if kinematically accessible) can be copiously produced at the LHC.
Their signatures are distinctly
different from the usual MSSM topologies. Depending on the masses of the
particles involved, either multi-jet final states with high sphericity and
large missing transverse momentum are predicted, or four top quarks should be
observed in $2\sigma$ production.  If the mass splitting between $L$ and $R$
squarks is not too small, loop--induced single $\sigma$ production may also
have a sizable cross section.
Though this channel suffers in general from large
backgrounds, identifying the $\sigma$ particle as a resonance in
2-gluon final states would truly be an exciting experimental observation.

\section*{Acknowledgments}

Supported in parts by the Korea Research Foundation Grant (KRF-2008-521-C00069), the German BMBF grant  (05HT6PDA), the EU Research Networks ``UniverseNet'' (MRTN-CT-2006-035863), ``ForcesUniverse'' (MRTN-CT-2004-005104), ``The Quest for Unification'' (MRTN-CT-2004-503369), and the EC Programme``Particle
Physics and Cosmology: the Interface'' (MTKD--CT--2005--029466). JK is grateful  to
the Theory Division for the hospitality extended to him at CERN, PMZ to the Insitute of Theoretical Physics E at RWTH Aachen.

\end{document}